\documentstyle[preprint,aps,prl]{revtex}

\begin{document}

\draft

\title{ Vortex excitation in superfluid $^{\bf 4}$He: \protect\\ A
Diffusion Monte Carlo study }

\author{ S. Giorgini$^{1,2}$, J. Boronat$^{2}$ and J. Casulleras$^{2}$ }

\address{$^{1}$Dipartimento di Fisica, Universit\`a di Trento,
\protect\\ and Istituto Nazionale di Fisica della Materia, I-38050 Povo,
Italy} \address{$^{2}$Departament de F\'{\i}sica i Enginyeria Nuclear,
Campus Nord B4--B5, \protect\\ Universitat Polit\`ecnica de Catalunya,
E--08028 Barcelona, Spain}


\maketitle

\begin{abstract}
We present a diffusion Monte Carlo study of a single vortex in
two-dimensional superfluid liquid $^4$He within the fixed node
approximation. We use both the Feynman phase and an improved phase which
includes backflow correlations to model the nodal surface of the vortex
wavefunction. Results for the particle density,
core radius and  excitation energies are presented.

\end{abstract}


\narrowtext

Diffusion and Green's function Monte Carlo simulations have become a
standard tool in the study of ground-state properties of Bose quantum
liquids at zero temperature.  However, the development of
{\it ab initio} computational methods to investigate properties of
excited states, such as the phonon-roton branch and vortex excitations
in superfluid $^{4}$He, is still a challenging problem at the forefront
of present research in the field of computational techniques applied to
condensed matter physics.  In this Letter we present results on the
microscopic structure of a single vortex excitation in two-dimensional
liquid $^{4}$He at zero temperature, obtained by employing a fixed-node
diffusion Monte Carlo (DMC) method \cite{RCAL}.

The idea that circulation in superfluid $^{4}$He is quantized is due to
Onsager \cite{O49}, whereas the possibility that vortices might have a
core of atomic dimensions was first put forward by Feynman \cite{F55}
who also proposed a microscopic wavefunction for the vortex state.
Gross, Pitaevskii and Fetter \cite{GPF} investigated the structure of
vortex states in a weakly interacting inhomogeneous Bose gas using a
mean field approach. The first attempt to study quantized vortices in a
strongly interacting system, such as liquid $^{4}$He, is due to Chester,
Metz and Reatto \cite{CMR} who calculated the energy of a vortex line
with the use
of a variational approach involving integral equations.  Only
quite recently have appeared new calculations of the structure of vortex
states in superfluid $^{4}$He \cite{OC95,D92,SK,VRCK}.
The very recent paper by Ortiz and Ceperley \cite{OC95} is the
first attempt to tackle the problem of the vortex core structure by
employing {\it ab initio} computational techniques.  Our method in the
present Letter is similar to that used by these authors, but the
approach is different and our results for the core energy and the
particle density near the vortex axis are significantly different from
the ones obtained in Ref. \cite{OC95}.

A vortex excitation is an eigenstate of the $N$-particle Hamiltonian $H$
and of
the $z$-component of the angular momentum $L_z$ with eigenvalue $\hbar N
\ell$, corresponding to an integer number $\ell$ of quanta of
circulation \cite{Don}.  The simplest microscopic wavefunction to
describe a vortex state was introduced by Feynman \cite{F55}:
\begin{equation}
\psi_F({\bf R}) = e^{i\ell\varphi_F} \prod_{i=1}^N f(r_i)\Phi_0({\bf R})
\;\;,
\label{fey}
\end{equation}
where $\varphi_F =
\sum_{i=1,N}\theta_i$ is the Feynman phase with $\theta_i$ the azimuthal
angle of the $i$-th particle, $\Phi_0({\bf R})$ describes the ground
state of the
system and $f(r_i)$ is a function of the radial distance of
each particle from the vortex axis, which models the density near the
core.  In what follows we only consider vortices with one quantum of
circulation, i.e., $\ell = \pm 1$.  In the recent paper by Ortiz and
Ceperley \cite{OC95}, a systematic method to improve the phase of the
wavefunction is devised.  Starting from the Feynman phase $\varphi_F$ as
zeroth order ansatz, the first correction includes backflow correlations
giving an improved phase of the form
\begin{equation}
\varphi_{BF} =
\varphi_F + \lambda\sum_{i\neq j}\gamma(r_i,r_j,r_{ij}) \frac{r_j}{r_i}
\sin(\theta_i-\theta_j)\;\;.
\label{bf}
\end{equation}
The wavefunction
constructed with the phase $\varphi_{BF}$ is the vortex analogue of the
Feynman-Cohen backflow wavefunction for the phonon-roton excitation
branch \cite{FC56}.

To go beyond a variational estimation of the properties of the vortex
state, given by the above model wavefunctions, we have used a DMC
method.
This method solves the many-body Schr\"odinger equation in imaginary
time for the function $f({\bf R},t) = \psi_T({\bf R})\Phi({\bf R},t)$
\begin{equation}
-\frac{\partial f({\bf R},t)}{\partial t} = -
D\nabla_{\bf R}^2 f({\bf R},t) + D\nabla_{\bf R}({\bf F}({\bf R})f({\bf
R},t)) + (E_L({\bf R}) - E) f({\bf R},t) \;\;,
\label{sch}
\end{equation}
where $\Phi({\bf R},t)$ is the wavefunction of the
system and $\psi_T({\bf R})$ is a trial function used for importance
sampling.  In the above equation, $E_L({\bf R})=\psi_T^{-1}({\bf
R})H\psi_T({\bf R})$ is the local energy and ${\bf F}({\bf
R})=2\psi_T^{-1}({\bf R})\nabla_{\bf R} \psi_T({\bf R})$ is the
so-called quantum drift force; $D=\hbar^2/2m$, with $m$ the mass of the
particles, plays the role of a diffusion constant, ${\bf R}$ stands for
the $3N$-coordinate vector of the $N$ particles of the system and $E$ is
an arbitrary energy shift.  Equation (\ref{sch}) is a diffusion
equation for the probability distribution $f({\bf R},t)$ which evolves
in time due to diffusion, drift and branching processes.  If $\Phi({\bf
R})$ represents the wavefunction of the lowest energy eigenstate of the
system not orthogonal to the trial function $\psi_T$, the asymptotic
solution of Eq.  (\ref{sch}) is given by $f({\bf
R},t\rightarrow\infty)=\psi_T({\bf R})\Phi ({\bf R})$ and the
corresponding energy eigenvalue can be calculated exactly.

In order to deal with a real walker probability distribution function
$f({\bf R},t)$, we choose as trial function the superposition of two
vortex states, one with positive and one with negative circulation,
which are degenerate in energy.  The importance sampling function has
therefore the form
\begin{equation}
\psi_T({\bf R}) = \cos(\varphi({\bf
R}))\prod_{i=1}^N f(r_i) \psi_T^0({\bf R}) \;\;,
\label{tri}
\end{equation}
where $\psi_T^0$ is the importance sampling function for
the ground state and for the phase $\varphi$ we have used both the
Feynman phase $\varphi_F$ and the backflow phase $\varphi_{BF}$.  The
sign problem associated with the use of the above trial wavefunction in
the DMC algorithm has been dealt with in the framework of the fixed-node
(FN) approximation \cite{RCAL}.  This approach, which has been
extensively used in
the calculation of ground-state properties of fermionic systems, yields
an upper bound to the energy eigenvalue \cite{RCAL}.  The quantum drift
force acting on each particle, as obtained from the trial wavefunction
(\ref{tri}), can be written as the sum ${\bf F}^i({\bf R})={\bf
F}_1^i({\bf R})+{\bf F}_2^i({\bf R})$. The first term in the sum is
independent of the phase $\varphi$ of the wavefunction, whereas the
second term contains the $\varphi$ dependence and has the form
\begin{equation}
{\bf
F}_2^i({\bf R}) = -2 \tan(\varphi)\nabla_i\varphi_i \;\;\;,
\label{dri}
\end{equation}
where $\varphi_i$ is the contribution of the $i$-th
particle to the collective phase, $\varphi=\sum_{i=1,N}\varphi_i$.  In
the same way, the local energy $E_L({\bf R})$ can be decomposed in the
sum of a phase independent term $E_{L1}$ and a phase dependent term
$E_{L2}$ which is given by
\begin{equation}
E_{L2}({\bf R}) =
D\sum_{i=1}^N \left( (\nabla_i\varphi_i)^2 + \tan(\varphi)
\nabla_i^2\varphi_i - {1\over 2} {\bf F}_1^i({\bf R})\cdot{\bf
F}_2^i({\bf R}) \right) \;\;\; .
\label{elo}
\end{equation}

Vortex states are characteristic of systems with rotational invariance,
but, at the same time, simulations must deal with a finite number of
particles.  A simple choice is to restrict the $N$ particles to be
inside a cylindric box with the vortex at the center and rigid boundary
conditions on the walls.  This is the geometry chosen for example in
Ref. \cite{OC95}.  By confining the particles some problems arise, such
as the choice of the confining potential and surface effects which can
be relevant if the box size is not large enough.  In the present work,
we have removed the confinement in order to keep
surface effects as small as possible.  An important point one needs to
solve when attempting to simulate a vortex excitation in an infinite
system, is the calculation of the collective phase $\varphi({\bf R})$.
For a very large system one expects that
the features of the vortex state near the core are weakly influenced by
the behaviour of particles which are far from the vortex axis.
It is thus reasonable to assume that, if the collective phase is
decomposed in the sum
\begin{equation}
\varphi({\bf R}) = \sum_{r_i\leq\bar{r}}
\varphi_i + \sum_{r_i>\bar{r}} \varphi_i \;\; ,
\label{pha}
\end{equation}
where the first term sums the contributions to the phase of all the
particles with distance from the vortex axis within the cutoff length
$\bar{r}$ and the second term gives the contribution coming from all the
other particles, the phase fluctuations in the second term are
irrelevant for the core structure of the vortex and it can be safely
approximated by its mean value $\sum_{r_i>\bar{r}}\varphi_i = 0$.  If
this prescription is employed in the calculation of the collective phase
entering the expressions (\ref{dri}) and (\ref{elo}) for the drift force
and the local energy, one expects that for a large enough cutoff length
$\bar{r}$ the properties of the vortex state near the axis are properly
simulated.  However, the straightforward interpretation of the
decomposition (\ref{pha}) would be of no practical use because the
collective phase would change discontinously by a large amount each
time a
particle exits or enters the region delimited by the cutoff $ \bar{r}$.
Instead, the procedure we have adopted is to use the decomposition
(\ref{pha})
as a way of tagging the particles that will contribute to the collective
phase for a long simulation run.  The dependence of the results on the
cutoff length have been studied and as will be discussed later no
appreciable changes are seen for values of $\bar{r}$ larger than
approximately $3\; \sigma$ ($\sigma = 2.556$ \AA).  It is worth noticing
that in the limit $\bar{r}\rightarrow 0$ the collective phase vanishes
and all the terms containing explicitly $\varphi$ in the expressions
(\ref{dri}) and (\ref{elo}) for the drift force and the local energy
disappear.  In this case, the DMC calculation can be shown to be
equivalent
to the fixed-phase (FP) method employed by Ortiz and Ceperley in Ref.
\cite{OC95}, where the phase of the wavefunction is fixed and the DMC
algorithm is used to solve the equation for the modulus of the
wavefunction.  By taking a finite value for the cutoff $\bar{r}$, one
allows for phase fluctuations in the system around the phase introduced
with the trial function, and in the limit $\bar{r}\rightarrow \infty$
one recovers the full FN approximation.  Our results actually show
that a finite cutoff is enough
to account  completely for the phase fluctuations.

Once established that the contribution of the more distant particles to
the collective phase is irrelevant, there is no a compelling reason for
not extending the
system using periodic boundary conditions. Surface effects exist also in
this case, in the sense that particles near the walls of the box ``see"
the artificial density perturbations associated with the image vortices
in the adjacent boxes.  However, these effects are negligible for a
reasonable size of the simulation cell and are definitely smaller than
the density oscillations induced by rigid walls.

We are now in a position to discuss our results.  We consider 64
particles in a two-dimensional box of length $L=15.0 \;\sigma$, which
corresponds approximately to the equilibrium density of the
two-dimensional homogeneous liquid \cite{US}.  The atoms
interact through the two-body HFD-B(HE) potential \cite{AZI}, which is a
revised version of the HFDHE2 Aziz potential. The ground-state
trial wavefunction we have chosen is the McMillan two body function
$\psi_T^0({\bf R})=\prod_{i<j}\exp(-b^5/2r_{ij}^5)$ with
$b=1.205 \;\sigma$ as in the ground-state
calculation of Ref. \cite{US}.  For the radial function $f(r)$, which
models the structure of the vortex core, we consider two different
options: $f_1(r) = 1 - e^{(-r/a)}$, and $f_2(r) = 1$. The first function
gives a density in the trial function which decreases to zero at the
vortex axis over a distance of order $a$, for which we take the value
$a=1\;$\AA, whereas the second one does not contain any parameter
associated with the vortex core.  For the backflow function $\gamma$
entering Eq.  (\ref{bf}) we have used the same functional form
$\gamma(r_i,r_j,r_{ij})=\exp(-\alpha (r_i^2+r_j^2)-\beta
r_{ij}^2)$ and the same values for the parameters $\alpha$, $\beta$ and
$\lambda$ as in Ref. \cite{OC95}.  We have performed the calculation
using three different trial wavefunctions: $\psi_T^{F1}$
and $\psi_T^{F2}$ which correspond to the Feynman phase $\varphi_F$ with
the radial terms $f_1$ and $f_2$ respectively, and $\psi_T^{BF}$
corresponding to the backflow phase $\varphi_{BF}$ with $f_2$.  In Fig.
1 we show results for the particle density $\rho(r)$ using
$\psi_T^{F1}$, $\psi_T^{F2}$ and $\psi_T^{BF}$.  These results have been
obtained by means of mixed estimators which are significantly biased
by the choice of the trial wavefunction.  As apparent from Fig. 1, the
behaviour near the core is strongly affected by the introduction or
not of the
radial term $f_1(r)$. In order to remove the influence of the trial
wavefunction in mixed estimators of coordinate operators one can use
pure estimators.  In the present work we have
employed the method presented in Ref. \cite{BC95}.  The results for the
pure density profiles do not depend anymore on the use or not of the
radial term $f_1(r)$ and the three cases converge to a single result.
The common pure profile is presented in Fig. 2,
where it clearly appears that a zero particle density is reached on the
vortex axis.  This result is in contrast with the prediction of a
significant non-zero particle density on the axis obtained with the
backflow phase in Ref. \cite{OC95}.  In our opinion the result of Ref.
\cite{OC95} can be influenced by the extrapolation technique used by
the authors to improve the mixed estimator result.  In fact, the
usual linear extrapolation technique, which is accurate to the
same order
as the one employed in Ref. \cite{OC95}, would change their result for
the particle density on the vortex axis in an amount comparable to their
prediction.

In Table I
are reported the energies per particle for the different trial
wavefunctions, together with the energy per particle $E_0/N$ in the
ground state.  The two results for the Feynman phase are almost equal,
whereas
in the case of the backflow phase the system is slightly more bounded
in accordance with the improvement of the nodal surface.
The cutoff length $\bar{r}$ for the
calculation of the collective phase has been taken as $\bar{r}=6
\;\sigma$. In the case
of the Feynman phase reducing the cutoff length does not give any change
in the results down to the FP limit $\bar{r}=0$.  For the backflow
phase, though, the FP energy is slightly less negative $E_{BF}^{FP}/N =
- 0.8136 \pm 0.0020\;$ K.

The excitation energy $E_v(r)$ of a vortex
inside a disk of radius $r$ is obtained as the difference between the
total energy of the disk with and without the vortex $E_v(r) = E(r) -
E_0(r)$.  For large distances from the vortex axis $E_v(r)$ is usually
decomposed in an hydrodynamic tail, which depends logarithmically on
$r$, and a core energy $E_c$: $E_v(r) = \pi\hbar^2\rho_0/m\;\log(r/\xi)
+ E_c$, where $\rho_0$ is the homogeneous density far from the axis and
$\xi$ is the vortex core radius.  In Fig. 3 we show the vortex
excitation energy $E_v(r)$ for the different choices of the trial
wavefunction. For distances $r \gtrsim 6$ \AA\ $E_v(r)$ shows the
expected hydrodynamic behaviour with a small negative shift
of the backflow
energy with respect to the Feynman one in agreement with
the total energy results shown in Table I. For small $r$'s, the
estimation
of $E_v(r)$ is not exact and exhibits the influence of the trial
wavefunction used. It is worth noticing the absence of spurious
oscillations at large $r$, present in the results of Ref.
\cite{VRCK}, which are induced by the confining rigid walls.

The core radius $\xi$ can be estimated, following Ref. \cite{OC95}, as
the position of the maximum in the azimuthal circulating current
$J_\theta(r)$. The radial dependence of $J_\theta(r)$ has been
estimated from the pure density profile using the
expression for the current in the FP approximation at the Feynman
level, $J_\theta(r)=\rho_p(r)/r$.
The value obtained is $\xi =2.10 \pm 0.20$ \AA\ which is in agreement
with the result reported in Ref. \cite{OC95}. By calculating the
hydrodynamic contribution to the total energy for our square simulation
box, we get the core energy $E_c^{F1}=1.23 \pm 0.25$ K,
$E_c^{F2}=1.18 \pm 0.26$ K and $E_c^{BF} =
1.00 \pm 0.26$ K for the Feynman and backflow phases, respectively.
These values coincide with the results obtained by a fit to $E_v(r)$ for
$r > 6$ \AA.
Our results for $E_c$  are significantly smaller from the
ones obtained in Ref. \cite{OC95} and the values of $E_c^F$ are close
to the variational results of Ref. \cite{SK} based on
the Feynman phase.

In conclusion, we have studied the structure of a vortex excitation in
two-dimensional superfluid $^4$He using a DMC method within the fixed
node
approximation. The collective phase of the vortex has been dealt
with  a
method that allows for the use of periodic boundary conditions, removing
spurious surface effects introduced by the use of confining geometries.
The fixed phase approximation is recovered as a limiting case in our
approach. The result for the density profile predicts a zero particle
density on the vortex axis for the two model phases used. On the
other hand, the inclusion
of backflow correlations in the phase gives a slightly smaller upper
bound for the excitation energy. Finally, we would mention that the
fixed node DMC method used in the present work
permits the study of other excited states in liquid $^4$He. Work is in
progress to extend our calculations to the
phonon-roton branch and the
vortex-antivortex excitation in two dimensions.

This work has been supported in part by DGICYT(Spain) Grant No.
PB92-0761 and No. TIC95-0429. S. G. acknowledges a post-doctoral grant
from the Ministerio de Educaci\'on y Ciencia (Spain). The supercomputing
facilities provided by CEPBA and CESCA are also acknowledged.

\begin{figure}
\caption{Mixed density profiles.
Solid line, with $\psi_T^{F1}$;
short-dashed line, with $\psi_T^{F2}$, and long-dashed line with
$\psi_T^{BF}$.}
\end{figure}

\begin{figure}
\caption{Pure density profile.}
\end{figure}

\begin{figure}
\caption{Radial dependence of the excitation energy.
Solid line, with $\psi_T^{F1}$;
short-dashed line, with $\psi_T^{F2}$, and long-dashed line with
$\psi_T^{BF}$.}
\end{figure}

\begin{table}

\caption{Results for the energies per particle.
$E_{F1}/N$, $E_{F2}/N$, and $E_{BF}/N$ correspond to the trial functions
$\psi_T^{F1}$, $\psi_T^{F2}$ and $\psi_T^{BF}$, respectively. $E_0/N$ is
the ground-state energy.}

\begin{tabular}{cccc} $E_{F1}/N (K)$ & $E_{F2}/N (K)$ & $E_{BF}/N (K)$ &
$E_0/N (K)$ \\ \tableline -0.8162(16) & -0.8171(18) &
-0.8199(18) & -0.8957(25) \\ \end{tabular}

\end{table}

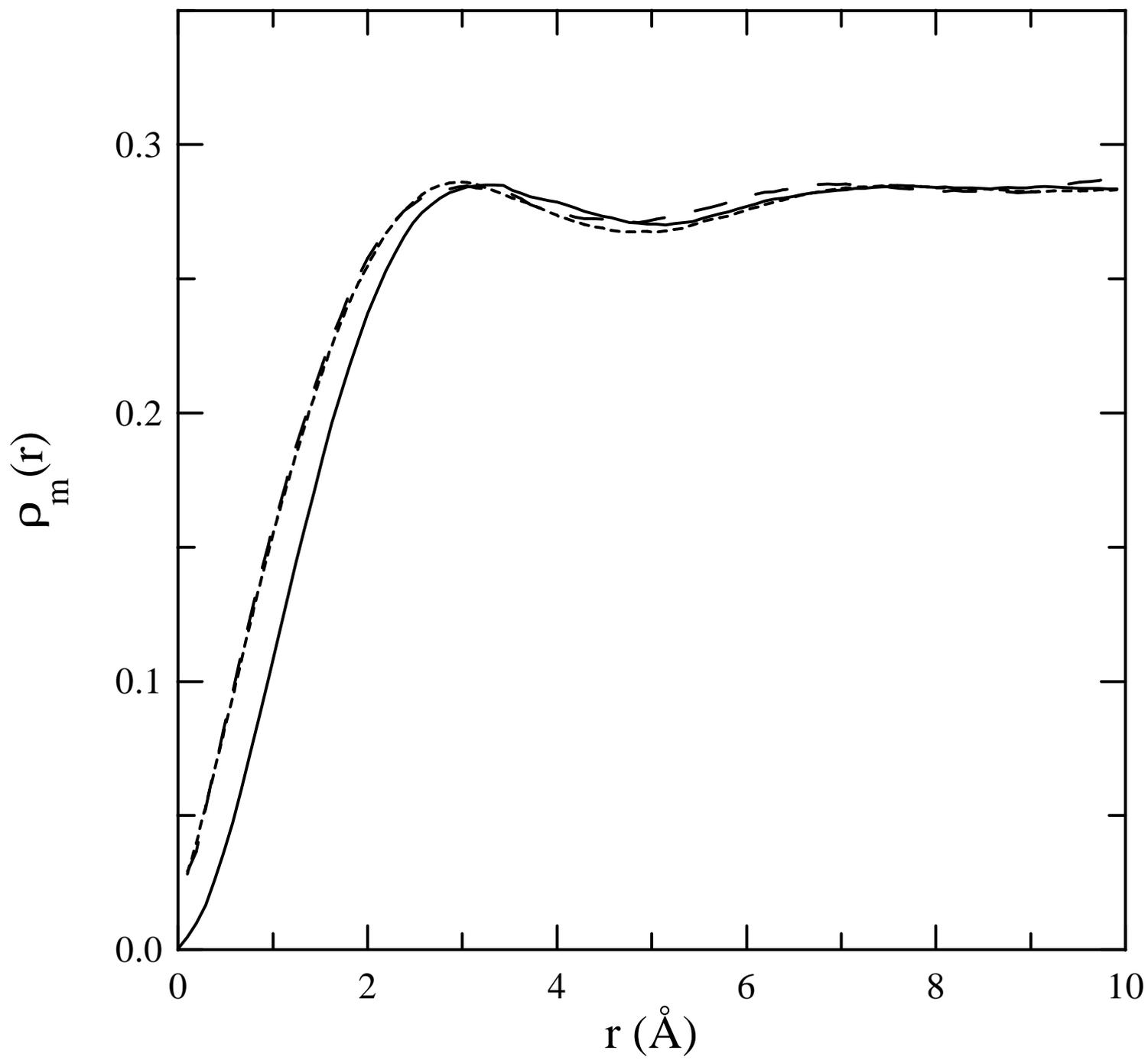
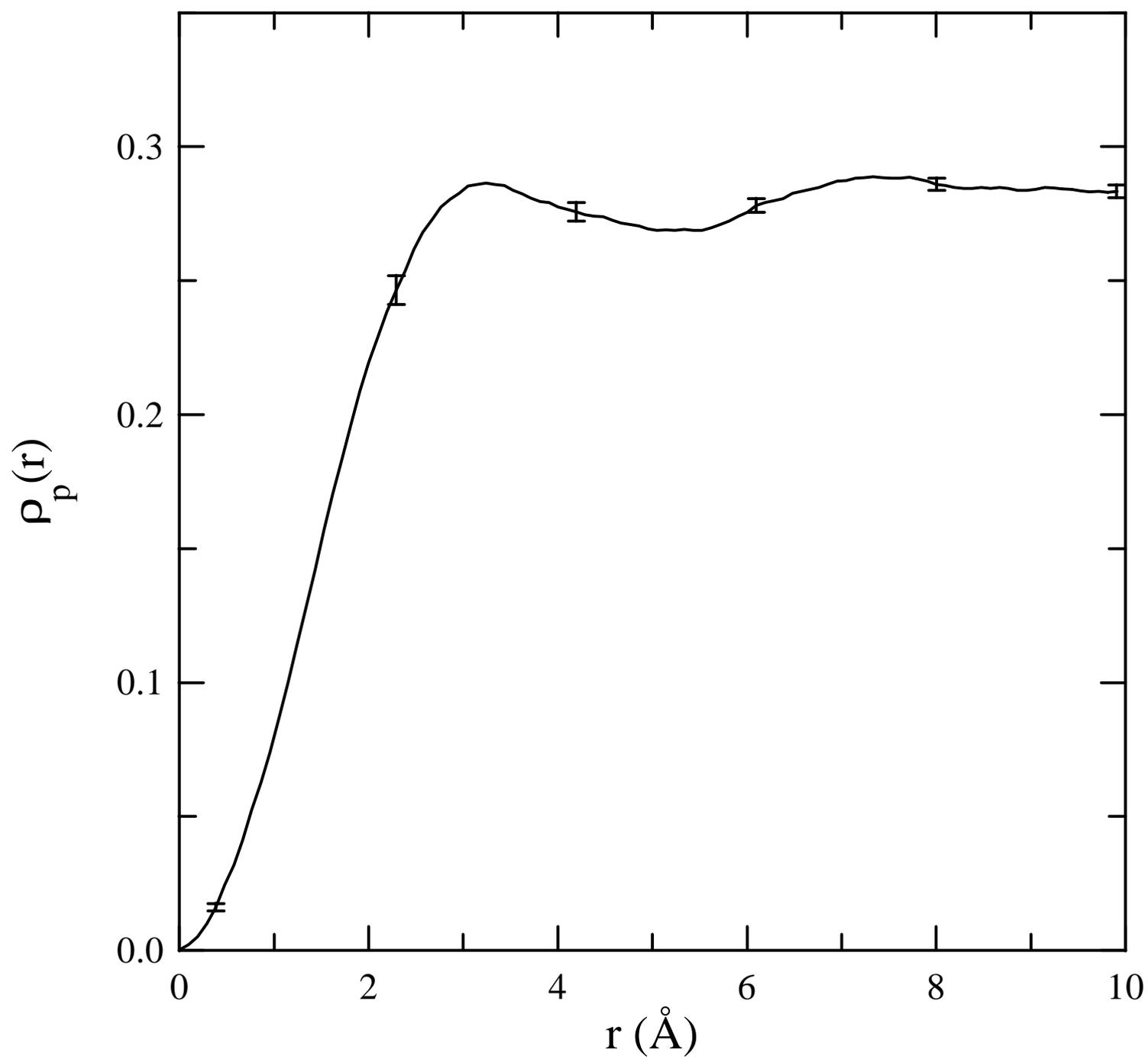
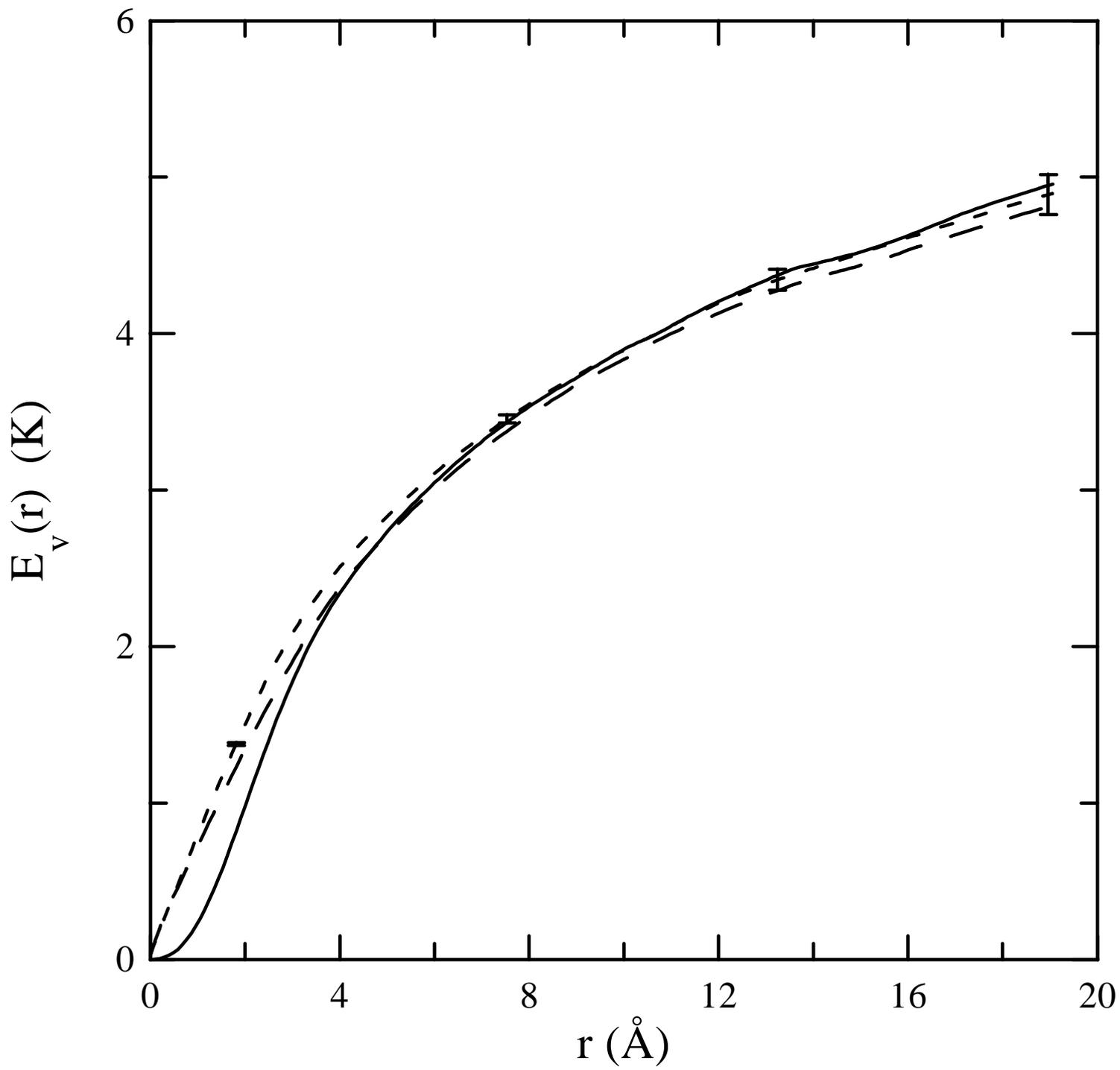
\end{document}